# Micromolar chemical imaging by high-energy low-photodamage Coherent Anti-stokes Raman Scattering (HELP-CARS)

Guangrui Ding[1,4], Dingcheng Sun[2,4], Yifan Zhu[1,4], Rong Tang[1,4], Hongli Ni[1,4], Yuhao Yuan[1,4], Haonan Lin[1,4*, #], Ji-Xin Cheng[1,2,3,4*]

[1] Department of Electrical and Computer Engineering, Boston University, Boston, MA, USA, 02215

[2] Department of Biomedical Engineering, Boston University, Boston, MA, USA, 02215

[3] Department of Chemistry, Boston University, Boston, MA, USA, 02215

[4] Photonics Center, Boston University, Boston, MA, USA, 02215

[#] New address: The Wallace H. Coulter Department of Biomedical Engineering, Georgia Institute of Technology and Emory University, Atlanta, GA, 30332

[*]Corresponding authors: haonan.lin@bme.gatech.edu; jxcheng@bu.edu

## Abstract

Coherent anti-Stokes Raman scattering (CARS) microscopy offers label-free chemical imaging capabilities, but its performance is constrained by small Raman scattering cross-section, strong non-resonant background (NRB), and limited signal-to-noise ratio (SNR). Here, we introduce a high-energy, low-photodamage CARS (HELP-CARS) platform designed to overcome these physical limitations. By employing a 1-MHz non-collinear optical parametric amplifier (NOPA) with extensive pulse chirping, HELP-CARS increases the coherent Raman excitation efficiency by ~300× and improves the signal-to-nonresonant background ratio by 11×, while inducing negligible damage during live cell imaging. Furthermore, to remove non-independent noise and physically entangled non-resonant background, we incorporate self-supervised deep-learning denoising and background removal based on the Kramers-Kronig relationship, yielding sensitivity improvement by an additional order of magnitude. Together, these advances provide a micromolar imaging sensitivity (160 $\mu$M for Dimethyl sulfoxide-d6) corresponding to 1000 molecules in the focal volume. Such high sensitivity enables high-fidelity chemical imaging in both fingerprint and silent windows. Hyperspectral HELP-CARS imaging of deuterated fatty acids allowed first observation of chemical separation with single lipid droplet. Together, HELP-CARS offers a powerful and generalizable approach for ultrasensitive and quantitative vibrational imaging of biological systems.

## Introduction

Optical microscopy [1] is pivotal to modern biological research and clinical practice. Quantitative label-free microscopy is especially powerful because of its ability to investigate biological objects in their native states, thereby circumventing fluorescent label's drawbacks including phototoxicity, photobleaching, and cellular function perturbations. Several label-free microscopy methods based on elastic scattering, such as holography[2,3] and diffraction tomographic imaging[4,5], are able to recover subcellular morphologies. These methods provide high-speed quantifications of optical phase delay or refractive index (RI) distributions with nanometer resolution and nanoscale sensitivity. As such, they are increasingly used in life science, such as living neuron activity evaluation[6], cell mass quantification[7], mitotic chromosome characterization[8], and volumetric tissue histopathology[9]. These imaging methods, however, are fundamentally limited by the lack of molecular specificity; thus, they are not able to resolve biochemical compositions of subcellular organelles or structures of biomolecules.

To achieve chemical specificity, nonlinear vibrational spectroscopic imaging techniques based on signals from intrinsic chemical bond vibrations have been developed[3,10] — for example, coherent Raman scattering microscopy has been developed for high-speed vibrational imaging of a broad range of biomedical systems[10-12]. In particular, the coherent anti-stokes Raman scattering (CARS)[13-15] microscope has demonstrated strong capacity in differentiating protein and lipid distribution in the biological samples[16]. However, the non-resonant background (NRB) intrinsic to the CARS process distorts the measured spectra and complicates the quantitative interpretation. By harnessing the spectrum causality, several numerical approaches -- such as maximum entropy method (MEM), Kramers-Kronig (K.K.) approach, wavelet prism (WP) analysis – have been developed to retrieve the true Raman response. More recently, deep-learning-assisted retrieval has improved reconstruction accuracy by incorporating these physical constraints[17].

Despite these algorithm advances, the fundamental sensitivity limits of CARS remain constrained by NRB-induced noise, especially when measuring low-concentration molecules. To overcome this physical limitation, polarization CARS (P-CARS), Time Resolved CARS (TR-CARS), Fourier Transformed CARS (FT-CARS) and Broadband-CARS (B-CARS) have been developed, but often at the cost of increased system complexity. Stimulated Raman Scattering (SRS) microscopy circumvents the NRB encountered in CARS and allows high speed, high fidelity hyperspectral chemical imaging with great success[18,19] . With these said, both CARS and SRS hyperspectral imaging requires tightly-focused laser beams with a large excitation power, resulting in a high potential for photodamage[20].

As a third order nonlinear process, the CARS intensity can be described as $I_{CARS} = \left|\chi^{(3)}\right|^2 I_P^2 I_S$. In the pulsed mode, the average power of CARS, pump and Stokes beams are related to pulse duration $\tau$ and repetition rate $f$ as $p_{CARS} = I_{CARS} * f * \tau$, $p_P = I_P * f * \tau$, and $p_S = I_S * f * \tau$. The CARS power is related to the pump and Stokes power and repetition rate as $p_{CARS} = |\chi^{(3)}|^2 \; p_P * p_P * p_S / (f * \tau)^2$ . Thus, by reducing the repetition rate from 80 MHz in OPO to 1 MHz in OPA, the CARS power can be enhanced by 6400 times under the same average pump and Stokes powers.

Inspired by the above calculation, we present a high-energy, low -photodamage CARS (HELP-CARS) imaging platform designed for super-sensitive and biosafe chemical imaging. HELP-CARS employs a 1-MHz non-collinear parametric oscillator (NOPA) to dramatically boost excitation peak power by ~80-fold for both pump and stokes fields (**Figure 1a**). The lower repetition rate of the NOPA enables much higher peak power at the same average power, which is essential for enhancing generation of the CARS signals. To address the photodamage issues, we chirp the femtosecond pump and Stokes pulses to 30 ps (pump) and 5 ps (Stokes), enabling live-cell imaging compatibility. These design features improve signal to NRB ratio by 11 times and yield a 4-fold better sensitivity over optical-parametric-oscillator (OPO) based SRS microscopy.

High-energy OPA excitation, however, introduces a spatially correlated and spectrally varying noise due to pulse to pulse instability and the stimulated Raman photothermal effect[21] . To address this issue, we incorporate a Self-Permutation Noise2noise Denoiser (SPEND)[22], a self-supervised deep learning framework tailored to remove non-independent and identically distributed (non-i.i.d.) noise. SPEND removes structured noise and improves the signal-to-noise ratio (SNR) of HELP-CARS by another one order of magnitude, reaching an unprecedented imaging sensitivity of 160 micromolar for DMSO-d6, corresponding to 1,000 molecules in the excitation volume. Together, the development of hardware and advanced denoising algorithms results in a 30-times better sensitivity HELP-CARS over OPO-SRS, enabling chemically specific imaging in silent and fingerprint windows.

# Results

## Theoretical calculation of signal and noise in OPO and NOPA based SRS and CARS

From previous work by Ozeki and co-workers[23], SRS and CARS signals at each pixel are expressed as:

$$S_{SRS} = \frac{2\epsilon}{\hbar\omega_p}\left(\frac{1}{2}\left|E_p\right|^2 - \frac{1}{2}\left|\mathrm{E_p} + \Delta E_{SRS}\right|^2\right)V_{eff} = \Gamma\gamma\mathrm{Im}(\chi_\mathrm{R})\left|\mathrm{E_p}\right|^2|E_s|^2 \tag{1}$$

$$S_{CARS} = \frac{2\epsilon}{\hbar\omega_{as}}|\Delta E_{CARS}|^2V_{eff} = \Gamma\gamma^2\omega_{as}|\chi_{NR} + \chi_R|^2\left|E_p\right|^4|E_s|^2 \tag{2}$$

Here, $\Gamma = \frac{2\epsilon c}{\hbar n}A_{eff}\Delta T f_{rep}\tau$, where $n$ is the refractive index,$c$ is the speed of light, $A_{eff}$ is the effective excitation area, $\Delta T$ is laser pulse duration, $f_{rep}$ is the reputation rate, and $\tau$ is the pixel dwell time. $\gamma = \frac{3l}{2nc}$, where $l$ is the effective interaction length. CARS benefits from the convolution of three fields and we assume a resolution enhancement for three directions as $\frac{\sqrt{2}}{\sqrt{3}} = 0.8$. Thus, the $A_{eff,CARS} = A_{eff,SRS} * 0.64$ and $l_{CARS} = l_{SRS} * 0.8$.

After substituting the laser repetition rate ($f_{rep,OPO} = 80\,MHz$, $f_{rep,NOPA} = 1\,MHz$), chirping conditions ($\Delta T_{Pump,OPO} = \Delta T_{Stokes,OPO} = 2\,ps$, $\Delta T_{Pump,NOPA} = 30\,ps$, $\Delta T_{Stokes,NOPA} = 5\,ps$) and peak power increasement ($|E_{NOPA}|^2 = 80 * |E_{OPO}|^2$) into the above expressions, we obtain that the NOPA-CARS photon yield exceeds that of OPO-SRS by a factor of ~172. Experimentally, we find an enhancement of ~300, in good agreement with the theoretical estimate.

In shot noise limited conditions, the noise of CARS and SRS (SRL) can be expressed as:

$$N_{SRS} = \sqrt{\frac{\Gamma\left|\mathrm{E_p}\right|^2}{\omega_\mathrm{p}}} \tag{3a}$$

$$N_{CARS} = \sqrt{S_{CARS}} \tag{3b}$$

To calculate the SNR of CARS, we define the vibrational contribution of the CARS signal as

$$V = \frac{|\chi_{NR} + \chi_R|^2 - |\chi_{NR}|^2}{|\chi_{NR} + \chi_R|^2} = \frac{2\chi_{NR} Im(\chi_R) + |\chi_R|^2}{|\chi_{NR} + \chi_R|^2} \quad (4)$$

In low concentrations, $|\chi_R| \ll \chi_{NR}$, and $V_{low} = \frac{2\chi_{NR} Im(\chi_R)}{|\chi_{NR}+\chi_R|^2} = \frac{2Im(\chi_R)}{|\chi_{NR}|}$,

$$S_{vib\ CARS} = V * S_{CARS} = 2\Gamma\gamma^2\omega_{as}\chi_{NR} Im(\chi_R)|E_p|^4|E_s|^2 \quad (5)$$

Converting the photon number to the power at the detector, we can get the SNR of SRS and CARS:

$$SNR_{SRS} = \left(\frac{S_{SRS}}{N_{SRS}}\right)^2 = \Gamma\gamma^2\ \omega_p|Im(\chi_R)|^2|E_p|^2|E_s|^4 \quad (6)$$

$$SNR_{CARS} = \left(\frac{S_{vib\ CARS}}{N_{CARS}}\right)^2 = \Gamma\gamma^2\omega_{as}\left(\frac{2\chi_{NR} Im(\chi_R)}{|\chi_{NR} + \chi_R|}\right)^2 |E_p|^4|E_s|^2 \quad (7)$$

The ratio of SNR for each modality reflects their respective sensitivity limits. For OPO excitation, The SNR of OPO-CARS is 0.66 of that of OPO-SRS, indicating that OPO-SRS and OPO-CARS sensitivity are of the same level. This result is consistent with the work by Ozeki et al.[23]. In practice, however, the OPO-CARS photon flux is too low to reach the detector's shot-noise limit, and detector noise dominates, resulting in poorer sensitivity than predicted.

For NOPA based CARS and SRS, the shot noise limited SNR ratios increase dramatically. Here, we assume $|E_p| = |E_s|$ and ignore the pulse chirping mismatch. The SNR of NOPA-CARS is 74 times of that of OPO-SRS. The SNR of NOPA SRS is 1024 times of that of OPO SRS, suggesting that sensitivity could, in principle, increase by approximately three orders of magnitude. Notably, these calculations are based on shot-noise-limit assumption, which does not necessarily hold true in experiments. Laser-instability-induced fluctuations dominate the noise floor, suppressing the achievable sensitivity. A detailed discussion of HELP-CARS system, non–shot-noise-limited behaviors, and a scheme for noise suppression are provided in the next sections.

### HELP-CARS system

We start with a comparison between OPO and NOPA (**Fig. 1a**). The lower repetition rate of NOPA laser enables higher peak power. The experimental setup of HELP-CARS is illustrated in **Fig. 1b**. A femtosecond NOPA produces 2 synchronized beams: a tunable pump beam ranging from 650 nm to 900 nm and from 1200 nm to 2500 nm, and a Stokes beam fixed at 1045 nm. For CARS excitation, we use a pump wavelength of 852 nm for the silent region and 893 nm for the fingerprint region. Hyperspectral imaging is achieved through spectral focusing: the pump pulses are temporally stretched to ~30 ps using 17 passes of 15-cm SF-57 glass rods, while the Stokes pulses are chirped to ~5 ps by 23 passes rods of the same type (**Fig.1c**). A motorized delay stage controls the temporal overlap between the two pulses.

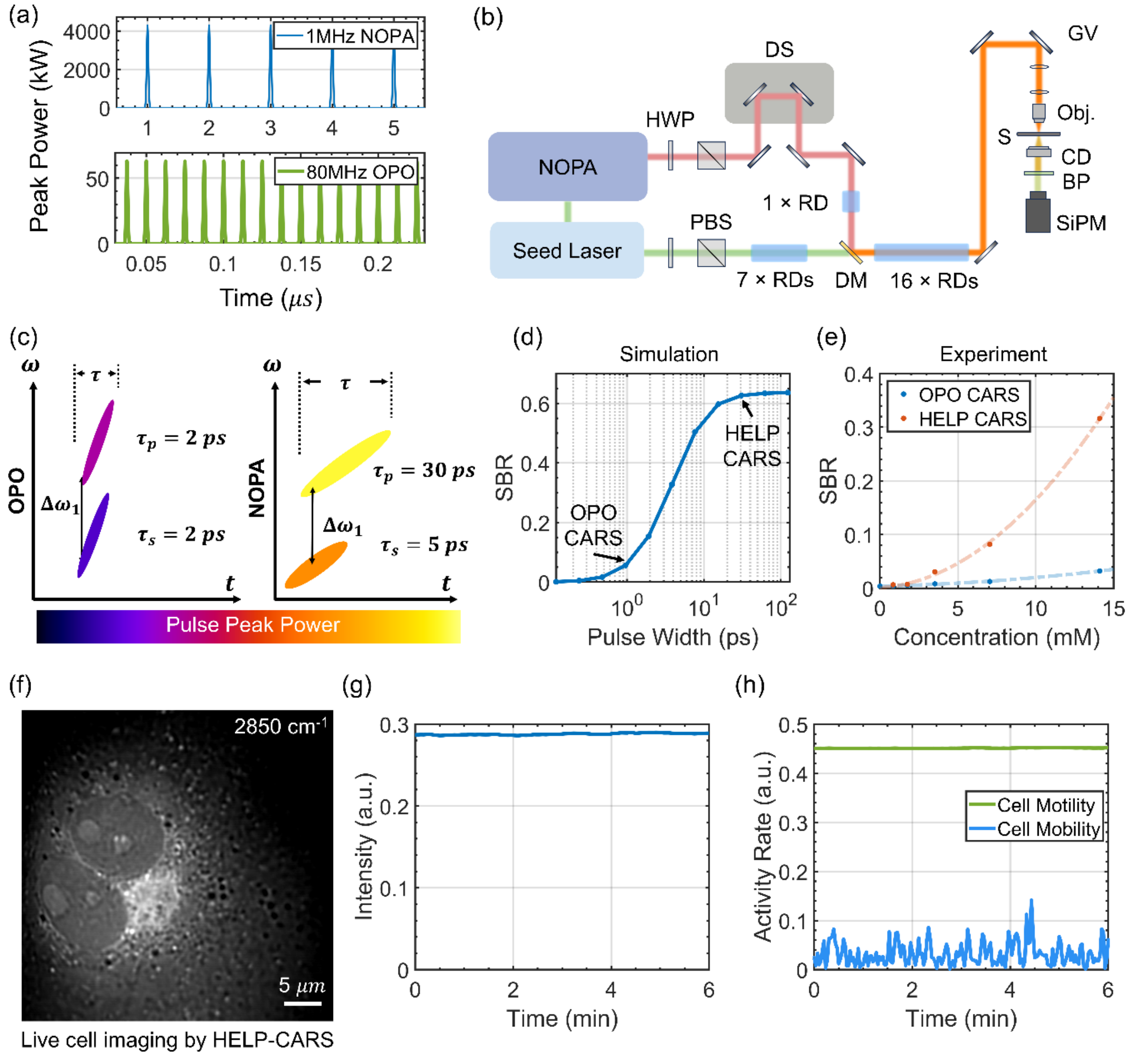

**Figure 1. Principal and schematic of HELP-CARS.** (a) Peak power comparison between NOPA and OPO excitation systems. (b) The experimental setup of HELP-CARS. HWP, half wave plate; DS, delay stage; RD, rod; PBS, polarizing beam splitter; DM, dichroic mirror; SU, scanning unit; O, objective; S, Sample; CD, condenser; BP, bandpass filter; SiPM, silicon photomultiplier. (c) Comparison of chirping conditions in NOPA and OPO based excitation. (d - e) Signal to NRB ratio from simulation (d) and experimental measurements (e). (f) Live-cell HELP-CARS imaging at 2850 $cm^{-1}$. (g) Average intensity of HELP-CARS imaging of living cells over 6 minutes duration. (h) Cell mobility and motility evaluation.

The CARS signal from 1% DMSO-d6 solution was measured to be ~1 $\mu$W at the detector. To avoid saturation, we choose a silicon photomultiplier (SiPM, Hamamatsu) rather than a photomultiplier tube (PMT). Prior work has established that appropriate chirping conditions can help to suppress NRB[24]. According to simulation, in HELP-CARS, the chosen chirping schemes improve the signal-to-NRB ratio by 11x (**Fig. 1d**) over OPO-CARS, consistent with experiment measurements from pure chemical standards (**Fig.1e**).

Photodamage is a major concern in non-linear optical microscopies for biomedical imaging. Previous studies[20,25,26] have shown that stretching femtosecond pulses into the picosecond regime converts nonlinear photodamage mechanisms into predominantly linear photothermal effect, which can be mitigated by faster scanning speed. The chirping used in HELP-CARS ensures that the scanning power remains within live-cell-safe limits. **Fig. 1f-h** demonstrates low phototoxicity during live-cell imaging in the C-H region. At a Raman resonance of 2850 $cm^{-1}$, cells remain viable over a continuous 6-minute scanning. There is no significant change in average intensity, mobility and motility of a living cell. We note that the pump power and stokes powers before the microscope were set at 70 mW and 50 mW, higher than the powers used for subsequent experiments. Moreover, the Raman cross-sections in the fingerprint and silent region are smaller than those in the high-wavenumber C-H vibration region, further reducing the phototoxic risk.

### Limit of detection comparison between OPA/OPO based CARS & SRS

In theoretical analysis, we derived the expected limits of detection (LoD) for OPO/NOPA based CARS and SRS. Here, we experimentally scrutinize those predictions using DMSO-d6 diluted in water as a testbed. All LoD measurements use the mean value of a 5×5-pixel area with a pixel dwell time of 10 $\mu$s.

In OPO-based systems, SRS achieves a limit of detection of 28 mM (0.2%) (**Fig. 2a**). In contrast, OPO-CARS performs worse because the large NRB introduces substantial noise that overwhelms the resonant CARS signals (**Fig. 2b**). Although applying the K.K. relation helps recover the distorted CARS spectrum, the insufficient SNR restricts reliable reconstruction to concentration above 140 mM (1%) (**Fig. 2c**). We used biosafe OPO-SRS powers (30mW pump, 150 mW stokes) in these measurements.

Next, we evaluated NOPA-based modalities. Since SRS measures the intensity modulation of original wavelength, laser amplitude fluctuation directly converts into the noise. Given that NOPA exhibits stronger laser instability, NOPA-SRS suffers from severe noises resulting in a LoD worse than 1 M (10%) (**Fig. 2d**). Pure solution spectrum is demonstrated in **Fig. S1**. Conversely, NOPA-CARS (HELP-CARS) demonstrate much better sensitivity (**Fig. 2e**). After K.K. retrieval (**Fig. 2f**), HELP–CARS reaches an LoD of 6.8 mM (0.05%), representing a substantial improvement over all other configurations tested. Balanced detection improves the SNR of OPA SRS by ~4× in pure DMSO (**Fig. 2g, h**) but remains insufficient for low-concentration detection. Moreover, because CARS offers higher spatial and axial resolution than SRS[27], the CARS excitation volume is roughly half of the SRS excitation volume, reducing the number of molecules required for detection. At the 6.8 mM threshold, HELP-CARS detects as few as ~40,000 molecules, whereas OPO-SRS detects approximately ~280,000 molecules. This 7-fold reduction in required molecular quantity highlights the enhanced sensitivity of HELP-CARS.

A summary of the four modalities is shown in **Fig. 2i**. In OPO-based platforms, the SRS outperforms CARS. At high concentrations, the SNR of SRS and CARS can be comparable; however, as the concentration decreases, their behaviors diverge. SRS scales linearly with concentration, whereas CARS scales quadratically, causing the CARS signal to drop more rapidly. Meanwhile, the dominant noise in SRS originates from the local oscillator (LO) -- the transmitted laser field -- which remains constant and independent of sample concentration. In contrast, in CARS at low concentrations, the vibrational signal becomes much weaker than the NRB, and the NRB itself sets the noise bottleneck, leading to significantly degraded sensitivity. Whereas in NOPA-based platforms, the trend reverses: CARS outperforms SRS. This inversion arises from fundamental differences in laser noise behavior.

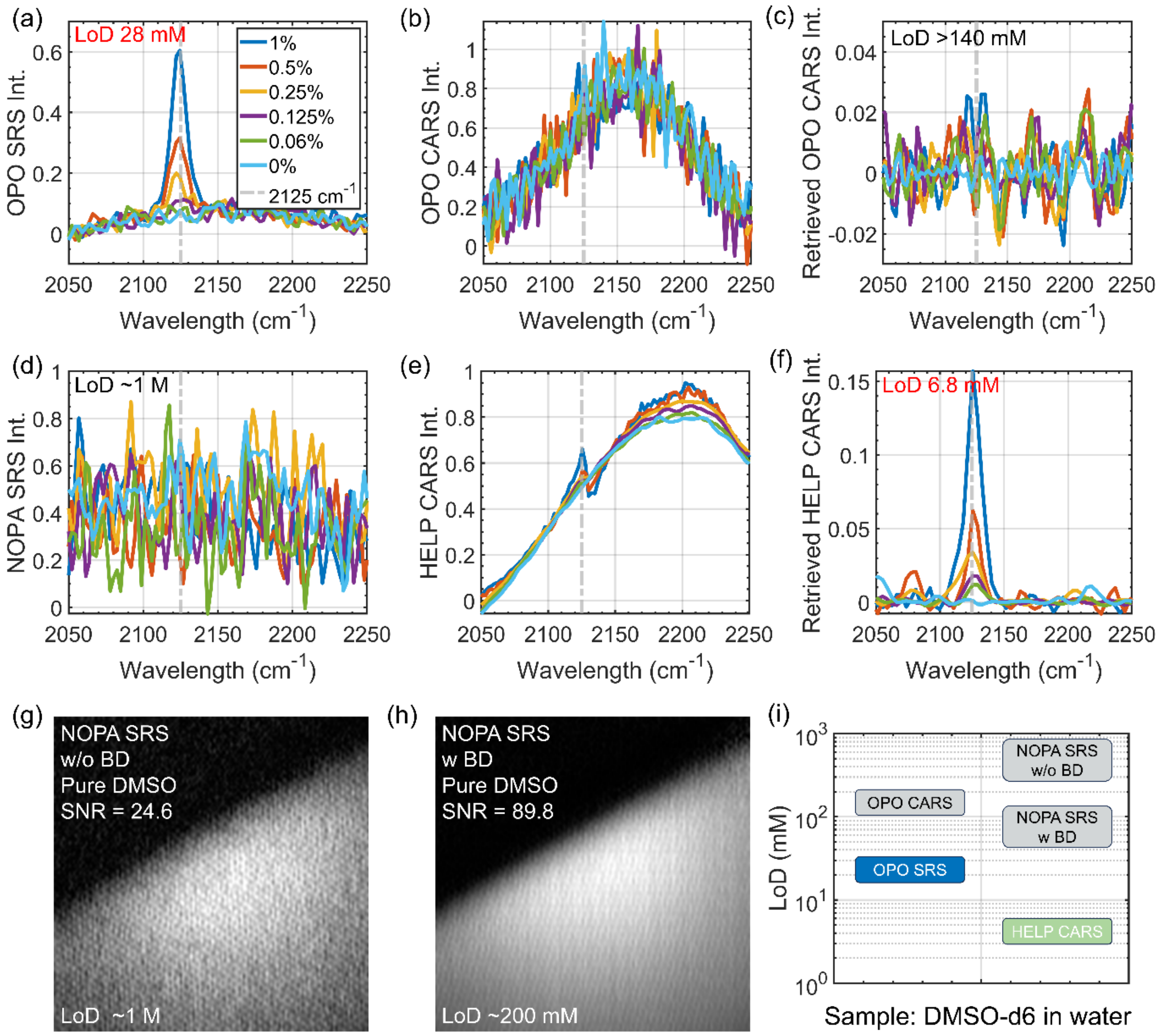

**Figure 2. Sensitivity comparison between OPO and NOPA based CARS and SRRS.** (a) Limit of detection (LoD) of OPO-SRS for DMSO-d6 diluted in water. (b) Raw OPO-CARS spectrum of DMSO-d6 in water. (c) LoD of OPO-CARS after K.K. retrieval. (d) LoD of NOPA-SRS for DMSO-d6 in water. (e) Raw NOPA-CARS (HELP-CARS) spectrum of DMSO-d6 in water. (c) LoD of HELP-CARS after K.K. retrieval. All LoD data are derived from the average of a 5×5-pixel region. (g – h) Effect of balanced detection on NOPA-SRS in pure DMSO: (g) without balanced detection; (h) with balanced detection. (i) Summary of sensitivity performance across OPO-SRS, OPO-CARS, NOPA-SRS, and NOPA-CARS (i.e., HELP-CARS).

We would provide a brief physical explanation of the cross-modality comparison between HELP CARS and OPO SRS. From the signal side, the higher peak power of the 1- MHz NOPA excitation boosts the coherent Raman process by 300 times compared to the OPO. From the noise side, in OPO SRS, the dominant noise is shot noise of the LO, which is a strong laser field. Milliwatt level LO power corresponds to $10^{16}$ photons, yielding shot noise on the order of $10^{8}$. In NOPA CARS, the LO is a non-resonant background, which is around microwatt level. The dominant noise is laser instability induced local oscillator fluctuation, experimentally measured at 0.1% - 1%, corresponding to fluctuations of $10^{9}$ – $10^{10}$ photons. This leads to 10 – 100 times higher noise in HELP CARS than that of OPO SRS. As a result, the increase of signal outweighs the increasement of noise in HELP-CARS, which means the SNR can be boosted over 3 times.

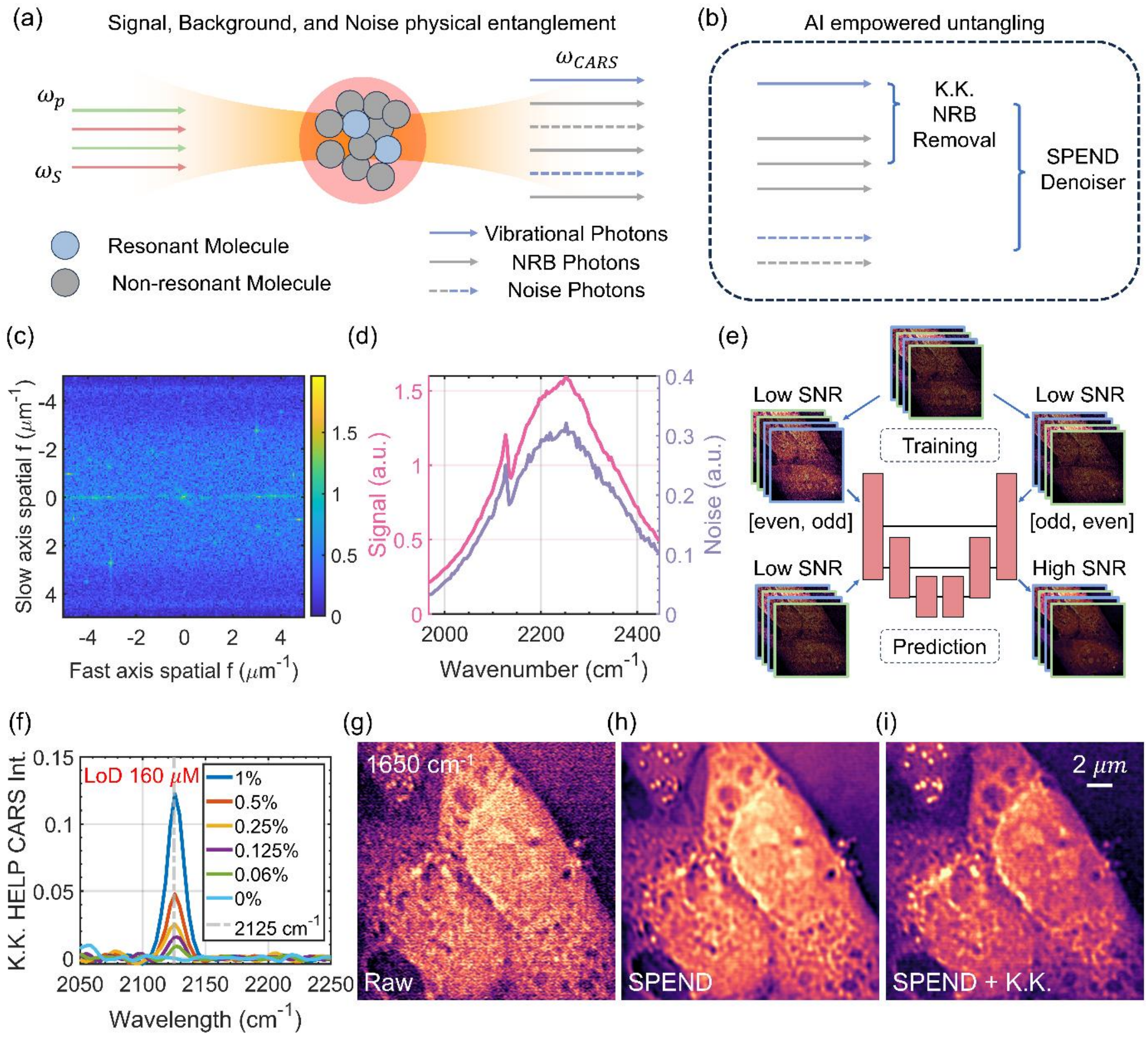

**Figure 3.** Data processing workflow and Sensitivity processing by SPEND. (a) Physical entanglement of signal, noise and background in CARS. (b) AI enables noise and background removal. (c) Spatial noise characterization of NOPA-CARS showing the power spectral density (PSD) along fast and slow scan axes. (d) Spectral variation of noise arising from the wavelength-dependent spectrum response. Sample: DMSO-d6. (e) Architecture of SPEND. (f) Sensitivity improvement of HELP-CARS after SPEND denoising. The data is from 5*5 pixels. (g - i) AI processing workflow for Cell data. (g) Raw cell imaging at 1650 $cm^{-1}$. (h) Cell image after SPEND. (i) Cell image after SPEND and background removal.

## Noise calibration and self-supervised deep learning denoising

As previously described, the HELP-CARS platform introduces higher noise levels than the OPO-SRS system, making it more reliant on effective post-processing denoising. Meanwhile, the nature of the nonlinear process inherently entangles the resonant signal, non-resonant background, and noise (**Fig. 3a**), resulting in a fundamental challenge for sensitivity enhancement. To address this, we leverage mathematical priors and artificial intelligence to disentangle these components and extract chemically meaningful information (**Fig. 3b**). Due to that noise directly degrades the accuracy of background-removal algorithms, denoising must be performed prior to NRB suppression. Previous work has shown that non-i.i.d. noise in OPO-SRS already challenges conventional denoising methods[22]. To fully characterize the noise behavior in HELP-CARS, we performed a detailed calibration of both spatial and spectral noise properties.

In spatial domain, high-energy excitation from NOPA introduces non-i.i.d. noise that violates the assumptions underlying classical and self-supervised denoisers. Slow amplitude fluctuations in the NOPA output produce slow-varying CARS photons, leading to spatially correlated noise across successive scan lines. Additionally, in spectral focusing based CARS experiment, the excitation profile is defined by the convolution of the pump, Stokes and probe fields, which forms an approximately Gaussian spectral envelope. As a result, the CARS photon fluctuation—not only the shot noise, which scales with the square root of photon number—varies across wavelengths. Moreover, the CARS measurement suffers from a vibrationally resonant thermal noise caused by the stimulated Raman photothermal effect. These effects produce structured spatial and spectral noise patterns that deviate strongly from the independent and identically distributed (i.i.d.) assumption.

To experimentally characterize these noise properties, we calibrated both spatial and spectral noise distributions in HELP-CARS system. As shown in **Fig. 3c**, the noise power spectral density (PSD) differs markedly between the fast and slow scan axes, confirming strong spatial noise correlation. Spectrally, the uneven excitation profile produces an uneven NRB distribution, resulting in wavenumber-dependent noise amplitudes (**Fig. 3d**). Together, these measurements verify the presence of non-i.i.d. noise consistent with the physical mechanisms described above.

Such noise patterns challenge conventional denoising algorithms. We compared three representative approaches—Noise2Void (N2V)[28], and Block matching and 4D filtering (BM4D)[29] and SPEND (**Fig. S2**). N2V underperforms because it assumes identical noise statistics across channels, leading to insufficient recovery of cellular structures under spectrally varying noise. BM4D struggles to suppress global laser-drift-induced fluctuations and fails to remove these correlated noise patterns. In contrast, SPEND (**Fig. 3e**), a self-supervised denoiser specifically designed for non i.i.d. noise in hyperspectral chemical imaging[22], performs robustly on HELP-CARS data. The strong NRB further facilitates the self-supervised training strategy to converge, enabling higher-fidelity reconstruction.

As a result, SPEND improves the sensitivity of HELP-CARS by an additional one order of magnitude, enabling detection down to 160 uM for DMSO-d6 in the silent window (**Fig. 3f**), corresponding to ~1000 molecules in the excitation volume. High fidelity cellular imaging at the fingerprint window is shown in **Fig.3 g-i**. SPEND improves the SNR by 11-fold. Phase retrieval by Kramer-Kronig (K.K.) relationship further removes the non-resonant background.

**HELP-CARS enables high fidelity chemical imaging of fatty acids in the silent window**

The silent window (2000–2300 $cm^{-1}$) offers a clean spectral window for bio-orthogonal imaging of alkyne- and deuterium-labeled biomolecules due to the absence of endogenous Raman signals. Through deuterium labeling in amino acids, glucose, cholesterol and fatty acids, SRS can be used to study protein, lipid synthesis, cholesterol storage and other biological processes[30]. However, the Raman cross-sections of these vibrational modes are typically small, making sensitive detection challenging in conventional CRS modalities. The improved excitation efficiency and high SNR of HELP-CARS opens new possibilities for robust imaging.

We evaluated the system performance using deuterium-labeled palmitate acid (PA-d31) incorporated into T24 cells. PA-d31 is a well-established indicator to directly visualize lipid processing in biological samples, which tracks the localization and metabolisms of fatty acids.

Single color SRS studies have visualized PA-d31 distributions, where the sensitivity is limited the low peak laser power and the broad cross-phase modulation (XPM) background [31-33].

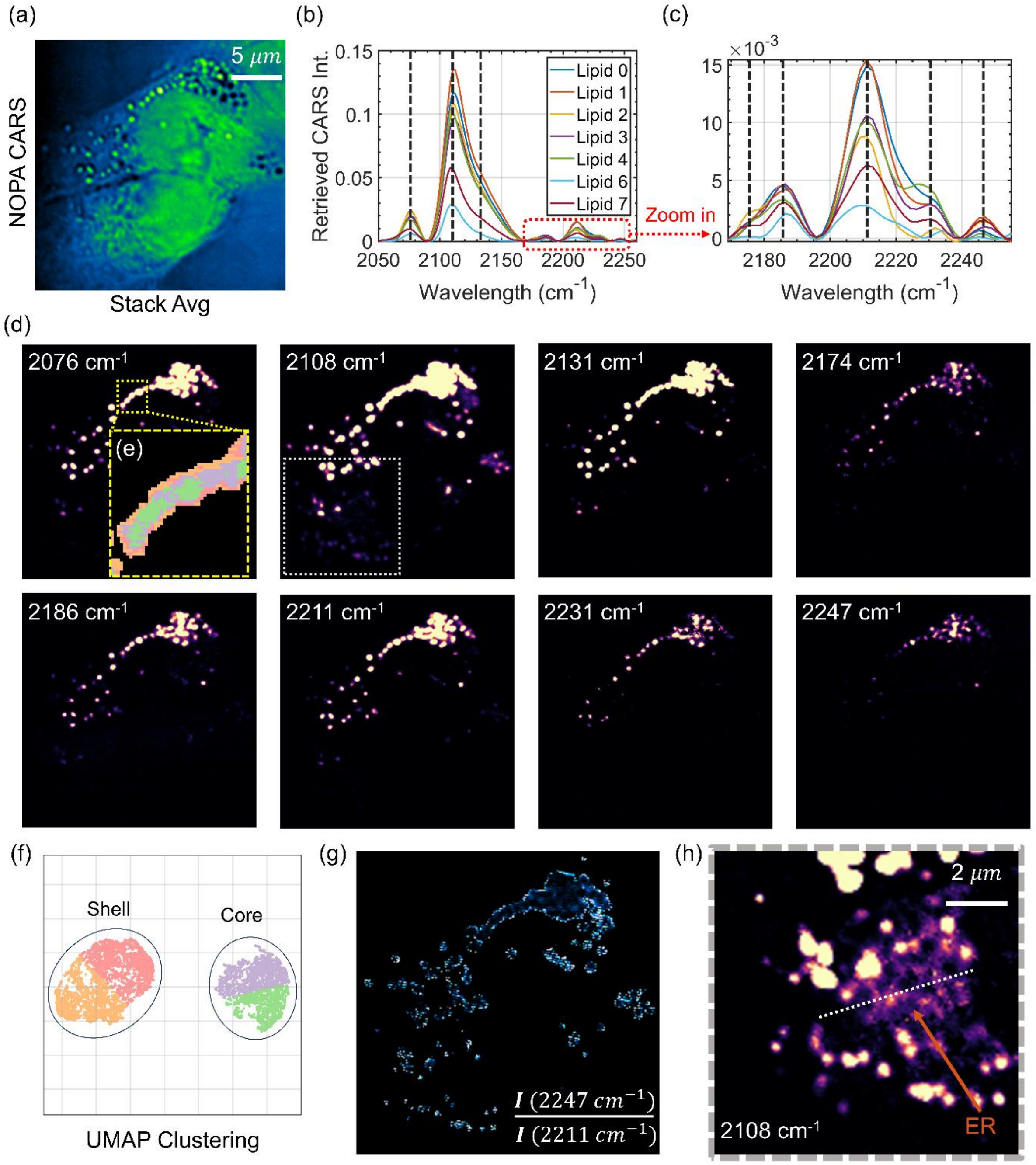

**Figure 4. HELP-CARS enables high fidelity imaging of fatty acids in the silent region.** (a) Average of HELP-CARS stack of PA-d31 labeled T24 cells. (b) Lipid CARS spectrum after K.K. retrieval (c) Zoom-in of selected spectrum window (d) Retrieved CARS map of eight characteristic wavelength. (e) Chemical clustering map. (f) UMAP clustering results. (g) Intensity ratio of 2247 $cm^{-1}$ and 2211 $cm^{-1}$. (h) Zoom-in of 2108 $cm^{-1}$ window.

Applying HELP-CARS, we obtain chemical maps of PA-d31 in the silent region (**Fig. 4a**). After phase retrieval, multiple characteristic C-D vibrational Raman modes are observed at 2076, 2108, 2131, 2174, 2186, 2211, 2231, and 2247 $cm^{-1}$, shown in **Fig 4b, c.** The selected lipid position is shown in **Fig. S3**. Peaks near 2076, 2108, 2211 $cm^{-1}$ corresponds to the $CD_3$[34], symmetric and asymmetric $CD_2$ vibrations[35]. The feature near 2131 $cm^{-1}$ has been reported as a shoulder of the 2108 $cm^{-1}$ peak and is associated with PA-d31 uptake rate[36]. Meanwhile, peaks at 2131, 2174, 2186 and 2231 $cm^{-1}$ have also been reported in deuterium-labeled proteins, nucleic acids and glycose[37-39], suggesting that these signals originate from PA-d31 metabolites. The 2247 $cm^{-1}$ peak represents D-C=C-D stretching, providing information about fatty acid saturation and desaturation[40].

Utilizing these eight spectral components as an eigenvector for each pixel, we cluster all pixels into four groups, as shown in **Fig. 4e**, revealing a clear core shell structure. After applying UMAP for dimension reduction[41], two lipid component sets emerge, corresponding to the shell and the core, respectively (**Fig. 4f**). Each further divided into 2 subsets. Averaged spectra retrieved from the masks (**Fig. S4**) show that the D-C=C-D peak intensity is significantly higher in the shell than in the core, indicating that desaturated lipids are more located in the shell. Such phase separation is further confirmed by overlay of the images at 2211 and 2047 $cm^{-1}$ (**Fig. 4g**). While LD's liquid to liquid crystalline phase separation was shown by polarization microscopy[42], our work shows for the first time the chemical separation with a single LD.

Importantly, symmetric $CD_2$ vibrations are observed in the endoplasmic reticulum (ER) membranes (**Fig. 4h**), where PA-d31 is rapidly desaturated into unsaturated fatty acids and shunted to lipid droplets for storage. The spectrum of different regions of interests further illustrates the enhanced sensitivity of HELP-CARS (**Fig. S5**). In the ER—where C–D concentration is low; the primary 2108 $cm^{-1}$ vibration is detected. The control group without PA-d31 treatment is shown in **Fig. S6**. After phase retrieval, no CARS signals were found for the characteristic peaks of CD bond (2076, 2108, 2210 $cm^{-1}$).

For comparison, we juxtapose the results from hyperspectral OPO SRS of the same sample (**Fig. S7**). Since XPM is much broader compared to the Raman peaks, we applied asymmetrically reweighted penalized least squares smoothing (arPLS)[43] to remove the background. After arPLS, OPO-SRS recovers the symmetric $CD_2$ and $CD_3$ peaks in lipid droplets but still fails to resolve the asymmetric $CD_2$ vibration and the rest 5 peaks, nor C–D signatures in the ER. These weak features are only accessible through HELP-CARS. Together, these results demonstrate that HELP-CARS provides high-fidelity, high-sensitivity imaging of deuterium-labeled fatty acids in the silent region, outperforming hyperspectral OPO-SRS and enabling detailed metabolic characterization at subcellular resolution.

**HELP-CARS enables high-fidelity chemical imaging of cells in the fingerprint window**

The coherent Raman signals in the carbon–hydrogen (C–H) stretching window (2800–3100 $cm^{-1}$) are spectrally congested, with broad and overlapping bands that limit chemical specificity in complex biological samples. In contrast, the fingerprint window provides a rich set of distinct, molecule-specific Raman peaks that allow clear differentiation of several biochemical components such as proteins, fatty acids, and cholesterol. However, the inherently weak Raman cross-sections in this region pose significant challenges for conventional CRS techniques. The enhanced sensitivity of HELP-CARS enables high-fidelity chemical imaging across this weak-signal spectral domain.

We focused on the 1650 $cm^{-1}$ Raman window, which contains contributions from the Amide I (proteins), acyl C=C (unsaturated fatty acid) and sterol C=C (cholesterol) bands. **Fig. 5a-c** illustrates the hyperspectral processing results for a hyperspectral CARS stack, from raw CARS acquisition to K.K. spectral retrieval and SPEND denoising. Benefiting from the higher z axial resolution of CARS and rich morphological contrast of the NRB, we observed fiber like structures inside nucleus (**Fig. 5d**). K.K. retrieval further enhances the contrast (**Fig. 5e**), revealing structures likely arising from chromatin organization or protein-rich nuclear scaffolds.

Spectra extracted from representative regions of interest (ROIs) are shown in **Fig. 5f**. Protein- and lipid-rich regions both show strong features near 1650 $cm^{-1}$, though the Amide I band is broader than the sharper acyl C=C peak in lipids. Cholesterol exhibits a distinct peak at ~1670 $cm^{-1}$, corresponding to the sterol C=C vibration. These spectral differences enable quantitative unmixing of the underlying chemical components.

We further applied Multivariate Curve Resolution (MCR)[44] to decompose the hyperspectral data into protein, unsaturated fatty acid, and cholesterol maps. The results are shown in **Fig. 5g**. We treated the sample with 100 μM Cholesterol or methyl-$\beta$-cyclodextrin (m$\beta$CD), a cholesterol-depleting agent. Statistical analyses (**Fig. 5h, i**) show that cholesterol treatment increases the abundance of both cholesterol and lipid, consistent with cholesterol uptake, esterification, and storage within lipid droplets. In contrast, m$\beta$CD treatment reduces both cholesterol and lipid signals. The m$\beta$CD extracts cholesterol via its hydrophobic cavity, increasing membrane tension and triggering compensatory lipolysis, thereby decreasing lipid content. The protein channel is used for calibration. Single-color images of the m$\beta$CD are shown in **Fig. S8**, and the control group in **Fig. S9**. Together, these results demonstrate that HELP-CARS provides high-sensitivity, chemically specific imaging in the fingerprint window, enabling quantitative analysis of protein, lipid, and cholesterol distributions with subcellular precision.

## Discussion

HELP-CARS is a high pulse energy, low-photodamage coherent anti-Stokes Raman scattering imaging platform that achieves a four-fold sensitivity improvement over SRS. When combined with SPEND, a self-supervised deep-learning denoiser tailored for non-i.i.d. noise, HELP-CARS gains an additional order-of-magnitude enhancement in effective sensitivity, reaching 160 μM for DMSO-d6. These advances enable high-fidelity, quantitative chemical imaging across both the fingerprint and silent vibrational windows—two spectral regions that have historically been difficult to access with high sensitivity.

In the silent window, HELP-CARS provided background free, quantitative chemical map of PA-d31, resolving $CD_3$, asymmetric and symmetric $CD_2$ vibrations with subcellular specificity. Hyperspectral HELP-CARS imaging of deuterated fatty acids allowed the observation of chemical separation of core-shell structure within a single lipid droplet. The ability to detect and map deuterated fatty acids in ER membranes demonstrates the system's enhanced sensitivity and spectral fidelity compared with conventional SRS and CARS platforms. In the fingerprint region, HELP-CARS enables unmixing of proteins, lipids, and cholesterol, validated through pharmacological perturbations using cholesterol rich and m$\beta$CD -mediated depletion. These results confirm both the chemical accuracy and biological relevance of hyperspectral HELP-CARS imaging.

Several advances can be made to further enhance the performance of HELP-CARS. Frequency-modulation CARS[45] and polarization-resolved CARS can provide additional background suppression and sensitivity gains, potentially pushing detection limits into the tens of micromolar range. Incorporation of lock-in amplifier may enable sensitive CARS imaging at ambient conditions[46]. These methods could synergize with HELP-CARS to overcome the detector saturation limits and unlock even higher sensitivity.

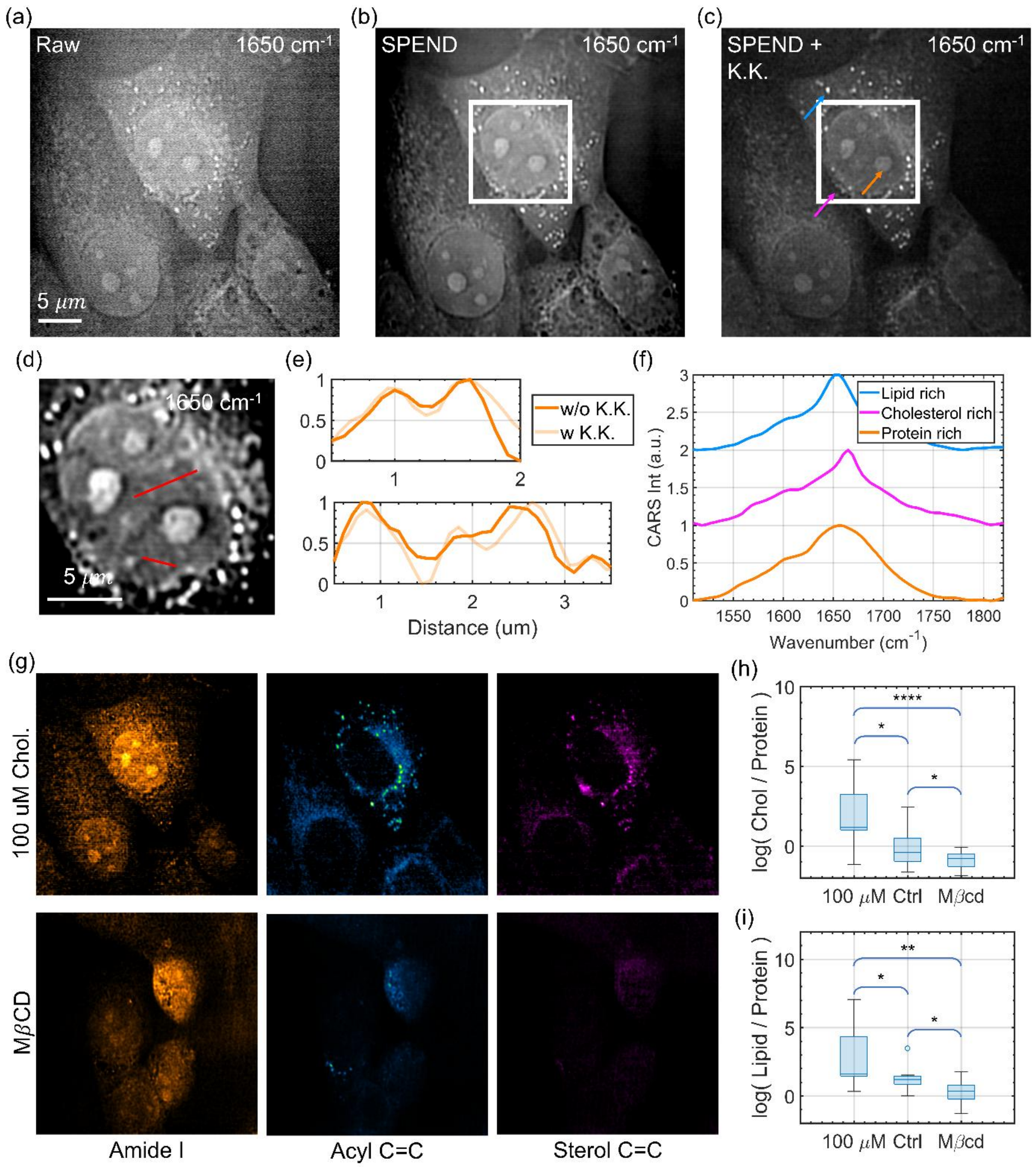

**Figure 5. HELP-CARS enables high fidelity chemical imaging in the fingerprint window.** (a) Raw HELP-CARS image at 1650 cm-1. (b) Denoised image at 1650 cm-1 using SPEND. (c) SPEND followed by K.K. retrieval at 1650 cm-1. (d) Zoomed-in view of the boxed region in (b & c). (e) Line profiles extracted before and after K.K. retrieval. (f) Representative K.K. retrieved HELP-CARS spectrum from lipid-rich, cholesterol-rich and protein rich regions. (g) Chemical unmixing map of protein (Amide I), Lipid (acyl C=C), and cholesterol (sterol C=C) for cells treated with 100 $\mu M$ cholesterol (top row) and methyl-$\beta$-cyclodextrin (m$\beta$CD) for cholesterol depletion (bottom row). (h) Quantitative comparison of cholesterol across treatment conditions. The results are normalized by protein channel. p (100$\mu$M, Ctrl) = 0.016. p (100$\mu$M, M$\beta$cd) = 5.82e-4. p (Ctrl, m$\beta$cd) = 0.043. (i) Quantitative comparison of fatty acid across treatment conditions. The results are normalized by protein channel. p (100$\mu$M, Ctrl) = 0.040. p (100$\mu$M, M$\beta$cd) = 0.003. p (Ctrl, m$\beta$cd) = 0.039.

HELP-CARS offers several promising applications for biological research. Micromolar-level sensitivity enables probing drug responses and metabolic rewiring in the fingerprint window, where chemically specific vibrational bands reside. In the silent region, HELP-CARS offers a powerful alternative to bio-orthogonal SRS without the drawbacks of cross-phase modulation. Finally, the strong coherent signal generation and NRB-based LO make HELP-CARS well-suited for epi-mode detection[15], suggesting exciting potential for in vivo or deep-tissue imaging applications.

More broadly, OPA-based excitation represents a powerful direction for next-generation chemical imaging. The ability to generate tunable, high-energy pulses provides a unique opportunity to balance peak intensity, spectral feasibility and photodamage. Alternative platforms, including wide-field OPA-CARS with random illumination, demonstrate the potential for high-throughput chemical imaging[47]. In this geometry, excitation energy is distributed across a large field of view, leading to a fundamental trade-off between throughput, per-pixel sensitivity, and spectral specificity. Together, these approaches highlight the versatility of OPA excitation in spanning complementary operating regimes, from high-sensitivity point-scanning microscopy to rapid wide-field chemical imaging.

## Methods

### HELP-CARS microscope

A lab-built hyperspectral Coherent Anti-stokes Raman microscope is used to perform hyperspectral CARS imaging. A femtosecond pulse laser (NOPA, spectra-Physics) operating at 980kHz with two synchronized beams, a tunable pump beam ranging from 650 nm to 900 nm, a fixed stokes beam at 1045 nm. The pump beam is tuned to 800nm for the C-H region, to 852nm for the C-D bond in the silent region, and 892nm for the fingerprint region. The pump beam is chirped using one 15-cm glass rod (SF57, Vibronix), while the Stokes beam is chirped by 7 passes of glass rods of the same type before the merging of two beams. The combined two beams were chirped by 16 passes of glass rods to picosecond pulse. The pump is then stretched to 30 ps and stokes is chirped to 5 ps. A motorized linear stage is used to tune the time delay between the pump and Stokes pulse which corresponds to the Raman shift of chemical bonds. A 2D Galvo scanner (GVS102, Thorlabs) is used for laser scanning. The combined beam is sent to the sample through a 60X water immersion objective (NA=1.2, UPlanApo/IR, Olympus). After interacting with the sample, the beam is collected by an oil condenser (NA=1.4, U-AAC, Olympus). A bandpass filter is applied to get rid of excitation field and purify CARS signals. We chose 650/45 nm for C-H region, 725/140 nm for silent region and 780/20 nm for fingerprint region. A SiPM (C13366-3050GA, Hamamatsu) is used to collect signals after filtering the Stokes beam.

### Photodamage quantification metrics

Cell motility and mobility are used to evaluate the cell migration capabilities to demonstrate the photodamage towards cells. Cell motility is defined as frame-to-frame intensity differences, which reflects intracellular dynamic activity

$$M_{motility}(t) = \frac{1}{N_{pixels}} \Sigma_{\mathrm{i \in Cell\ Mask}} |\mathrm{I_t(i)} - \mathrm{I_{t-1}(i)}|$$

Cell mobility is defined as the square root of mean square displacement (MSD)[48] of mass of center in chemical channel, which reflects cellular level migration behaviors

$$M_{mobility}(t) = \sqrt{|((r_{x,c}(t) - r_{x,c}(t-1))^2 + ((r_{y,c}(t) - r_{y,c}(t-1))^2|}$$

**Parallel Kramers-Kronig (K.K.) relation for NRB removal in hyperspectral HELP-CARS**

Kramers–Kronig (K.K.)–based CARS spectral retrieval was performed following previously reported algorithms[49]. Initial optimization of retrieval hyperparameters was conducted on single pixel 1D spectra to determine the optimal baseline correction hyperparameter settings. After identifying the optimal configuration, the full hyperspectral HELP-CARS datasets were processed using MATLAB with Parallel Computing Toolbox acceleration.

For a typical 400×400×100 hyperspectral stack, the K.K. retrieval required approximately 50–60 seconds using 8 CPU cores. All analyses were performed on MATLAB R2023b.

**Noise spectral and spatial analysis**

The non i.i.d. noise analysis method is reported[22]. To analyze spectral variation, we measured standard deviation within a small area to quantify noise level. Then the average intensity of the same area was calculated to represent the signal level. We, then plotted the relationship between the noise and signal to elucidate their dependency. Spatial correlation analysis was conducted using videos of single-color HELP-CARS.

**SPEND training and interference**

To enhance the robustness of our model, we augment the datasets by 4-fold through flipping and rotating at 180 degrees. After augmentation, the training set is comprised of 36 stacks, with 10% for validation and 90% for training. Each stack contains 400*400 pixels, and 200 frames. We employed a 4-layer Unet architecture based on the CSBDeep framework[50]. The training was conducted on a commercial graphics processing unit (GPU, RTX 4090, Nvidia), taking 2 hours to complete. For interference, it will take 18 seconds to denoise an entire image stack.

**Multivariate Curve Resolution (MCR) for chemical unmixing**

In the paper, we utilized MCR[44] for chemical unmixing. The dimensions of the hyperspectral data, x, y, $\lambda$, dissected as $N_x$, $N_y$, $N_\lambda$. For MCR, we first reshape the 3D hyperspectral stack into a 2D matrix ($D \in R^{N_x N_y \times N_\lambda}$) by arranging the pixels in the raster order. Assume the number of interested chemical channels is $K$, a model is used to decompose the data matrix into the multiplication of concentration maps $C \in R^{N_x N_y \times K}$ and spectral profiles of pure chemicals $S \in R^{K \times N_\lambda}$:

$$D = CS^T + E \qquad (1)$$

where E is the error. MCR-ALS is an algorithm that solves the bilinear model using a constrained Alternating Least Square algorithm, which improves the interpretability of the profile in both $C$ and $S^T$. MCR was implemented by a Python library, pyMCR[44].

### Asymmetrically reweighted penalized least squares (arPLS) smoothing for XPM removal

The arPLS is a numerical baseline correction method[43]. The baseline is assumed to follow the general trend of x while remaining smooth, which can be formulated as a regularized least-squares problem:

$$R(z) = (x - z)^T (x - z) + \lambda z^T D^T D z \tag{3}$$

where D is the difference matrix and $\lambda$ is a smoothness parameter. Introducing a diagonal weight matrix $W$ modifies Eq. (3) into a penalized least-squares objective:

$$P(z) = (x - z)^T W (x - z) + \lambda z^T D^T D z \tag{4}$$

Setting the partial derivative $\frac{\partial P}{\partial z^T} = 0$ yields the closed-form baseline estimate:

$$z = (W + \lambda D^T D)^{-1} W x \tag{5}$$

The PLS algorithm iteratively updates weights by comparing each estimated baseline $z_i$ with corresponding signal $x_i$. To reduce noise interference and to prevent negative deviations from being overweight, arPLS introduces an asymmetric weighting mechanism:

$$w_i = \begin{cases} logistic(d_i, m, \sigma) = \dfrac{1}{1 + e^{\frac{2(d_i - (-m + 2\sigma))}{\sigma}}}, x_i > z_i \\ 1, x_i < z_i \end{cases} \tag{6}$$

where $d_i = x_i - z_i$, and $m, \sigma$ are the mean and standard deviation of the negative d region. The algorithm iterates until convergence, that is when reaches the maximum number of iterations or weights change smaller than $\frac{|w_t - w_{t+1}|}{|w_t|} < r$, where $r$ is the ratio parameter, $w_t$ and $w_{t+1}$ are weights at t and t+1 iteration.

### Cell lines, chemicals, and cell culture PA-d31 treated T24 cells

T24 cells were purchased from the American Type Culture Collection (ATCC). T24 cells were cultured in high-glucose Dulbecco's modified Eagle's medium (DMEM, Gibco) supplemented with 10% fetal bovine serum (FBS, Gibco) and penicillin/streptomycin (P/S; 100 U/ml). All cells were maintained in a humidified incubator with a 5% CO2 supply at 37°C. Cells were seeded in 35 mm glass-bottom dishes for imaging experiments.

Palmitic acid-d31 (PA-d31, CAS Number: 39756-30-4) was obtained from Sigma-Aldrich. To feed the cell the isotope lipid, PA-d31 was initially dissolved in DMSO to the stock concentrations and subsequently diluted into culture medium. Cells were first seeded for 24 hours with fresh medium, followed by a 24-hour incubation with 100 μM PA-d31. To establish a cholesterol-loaded model, cells were treated with 100 μM cholesterol (CAS Number: 57-88-5) for 24 h before imaging. For cholesterol depletion, 100 μM methyl-β-cyclodextrin (MβCD; CAS Number: 128446-36-6) was added to the medium 2 h prior to imaging.

### Preparation of biological samples for imaging

For imaging of fixed cells, the cells were first washed with PBS (1×, pH 7.4, Thermo Fisher

Scientific Inc.) and then fixed with 10% neutral buffered formalin. This was followed by three subsequent washes with PBS. Unless otherwise specified, the culture medium was replaced with PBS immediately prior to imaging.

**Acknowledgements.**

The authors thank Prof. Lei Tian for discussion. This work is supported by NIH grants R35GM136223, R01EB032391, and R01EB035429 to JXC.

**Author contributions.**

G.D., H.L., and J.C. co-designed the project. D.S prepares biological samples. G.D., Y. Z., R.T., H. N., and Y.Y. conduct the experiment. H.L., and J.C. supervised the project. G.D., and J.C. wrote the manuscript. All authors read the manuscript.

**Data availability.**

The dataset and code have been made publicly available on https://github.com/buchenglab.

## Supplementary Information

# Micromolar chemical imaging by high-energy low-photodamage Coherent Anti-stokes Raman Scattering (HELP-CARS)

Guangrui Ding[1,4], Dingcheng Sun[2,4], Yifan Zhu[1,4], Rong Tang[1,4], Hongli Ni[1,4], Yuhao Yuan[1,4], Haonan Lin[1,4*,#], Ji-Xin Cheng[1,2,3,4*]

[1] Department of Electrical and Computer Engineering, Boston University, Boston, MA, USA, 02215

[2] Department of Biomedical Engineering, Boston University, Boston, MA, USA, 02215

[3] Department of Chemistry, Boston University, Boston, MA, USA, 02215

[4] Photonics Center, Boston University, Boston, MA, USA, 02215

[#] new address: The Wallace H. Coulter Department of Biomedical Engineering, Georgia Institute of Technology and Emory University, Atlanta, GA, 30332

[*]Corresponding authors: haonan.lin@bme.gatech.edu; jxcheng@bu.edu

This file contains Figure S1 to Figure S9.

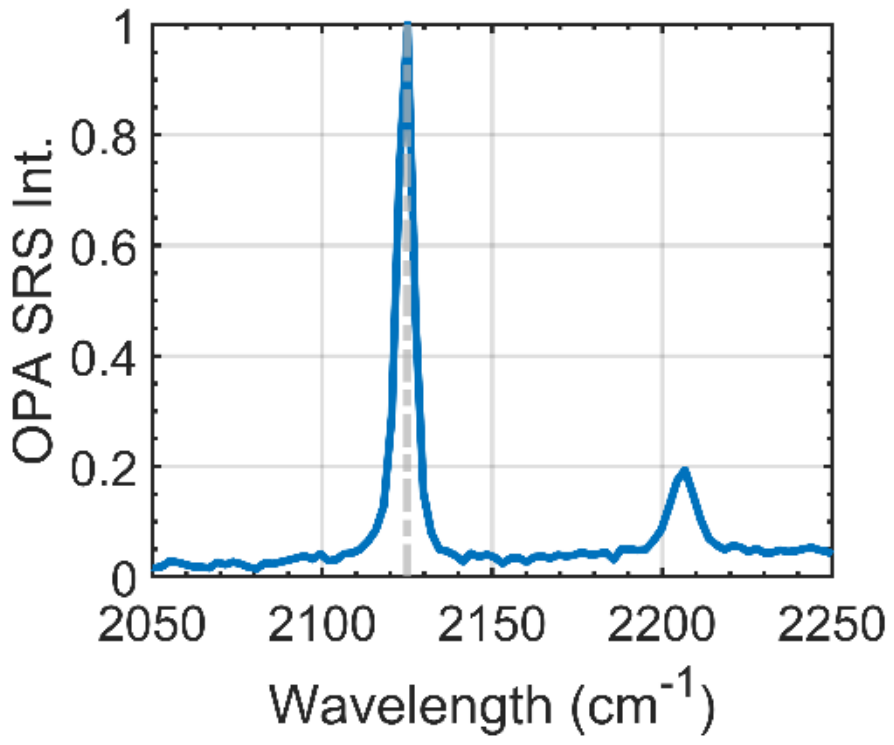

**Figure S1. NOPA SRS spectrum of pure DMSO-d6.**

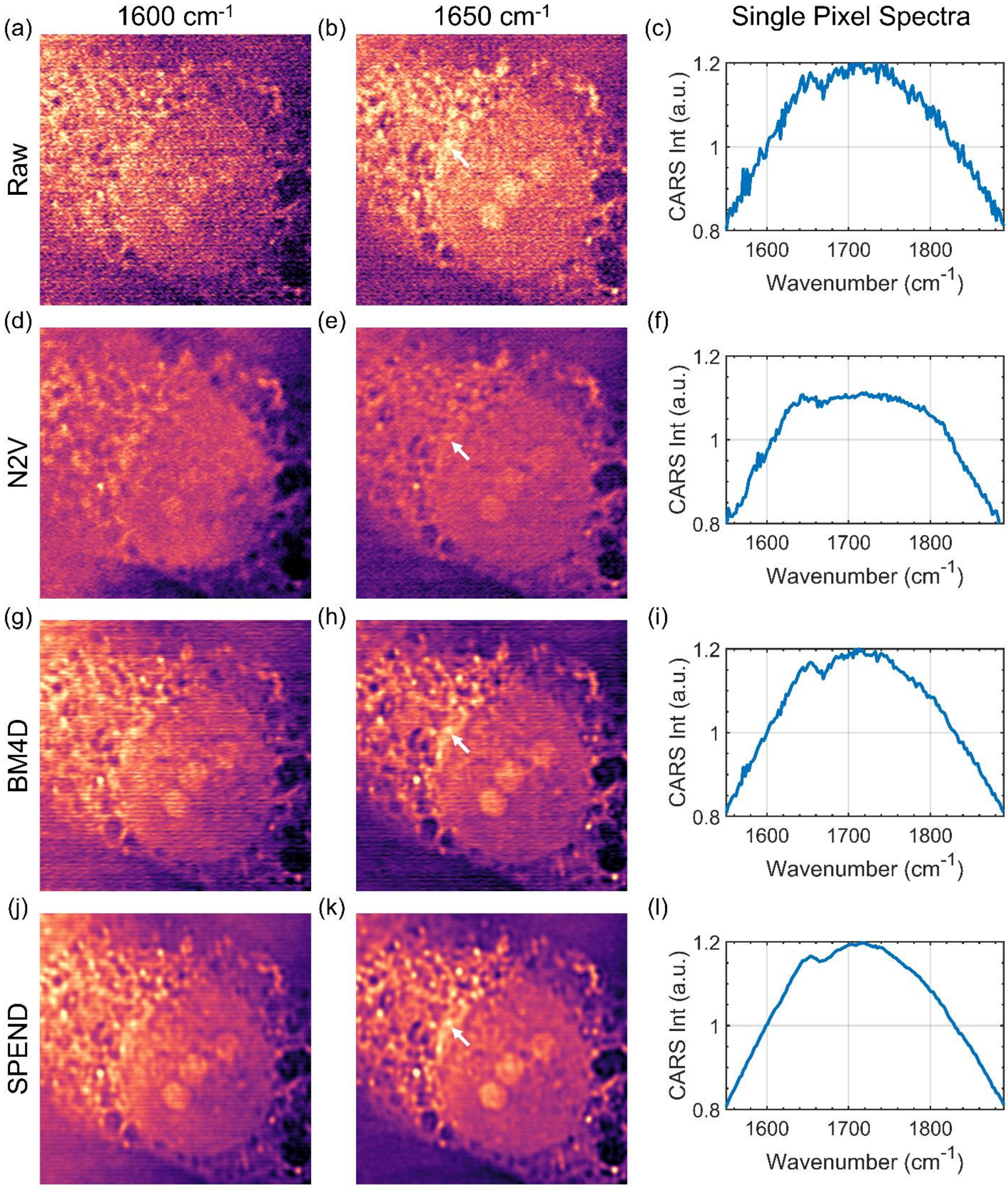

**Figure S2. Denoising Performance Comparison with BM4D/N2V/SPEND.** (a–c) Raw CARS images at 1600 $cm^{-1}$ (a) and 1650 $cm^{-1}$ (b), and corresponding single-pixel spectrum (c). (d–f) Noise2Void (N2V) denoising results at 1600 $cm^{-1}$ (d) and 1650 $cm^{-1}$ (e), with the single-pixel spectrum (f). (g–i) BM4D denoising at 1600 $cm^{-1}$ (g) and 1650 $cm^{-1}$ (h), and corresponding spectrum (i). (j–l) SPEND denoising results at 1600 $cm^{-1}$ (j) and 1650 $cm^{-1}$ (k), with the corresponding spectrum (l).

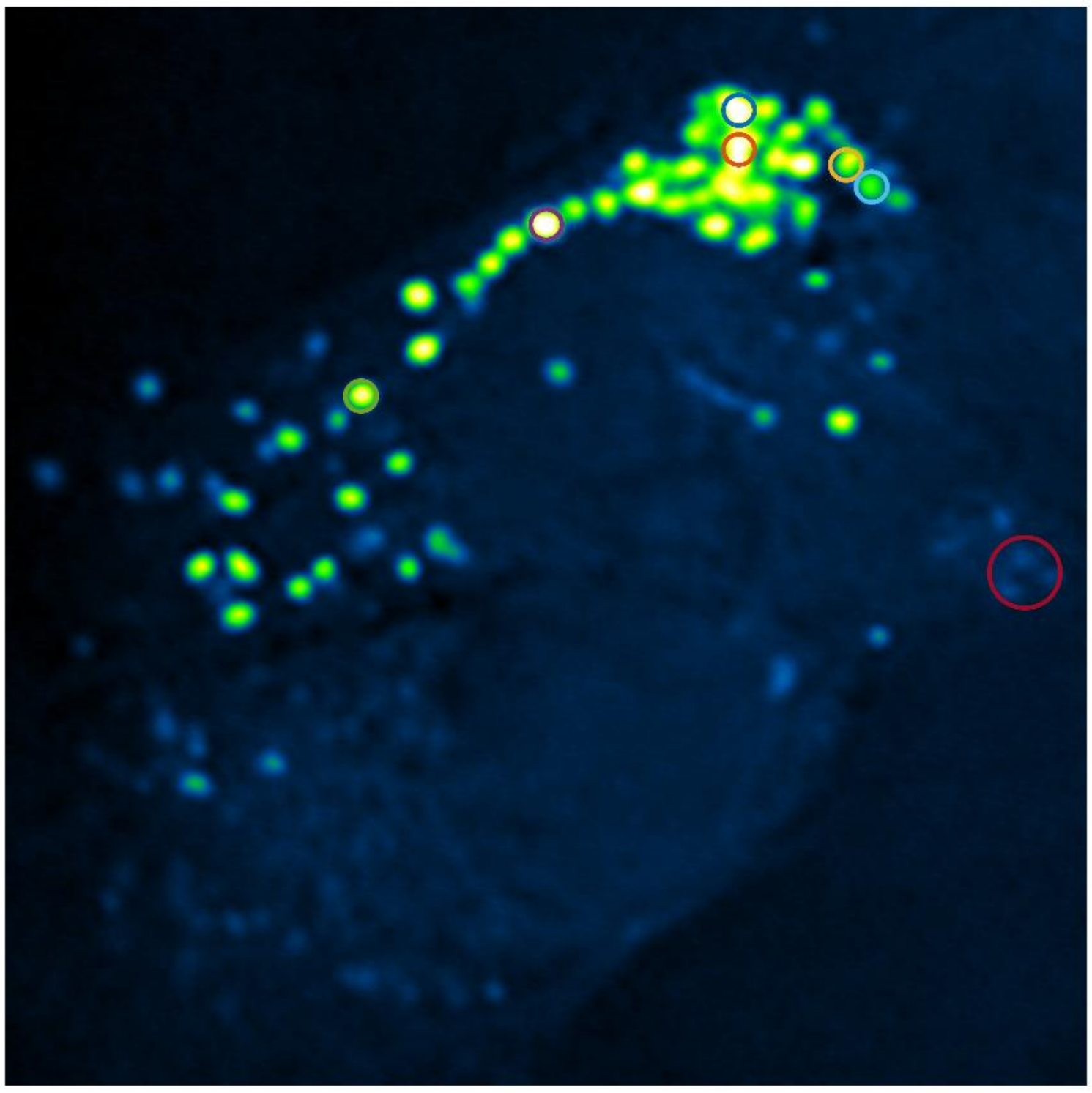

**Figure S3. Lipid droplet selection map.**

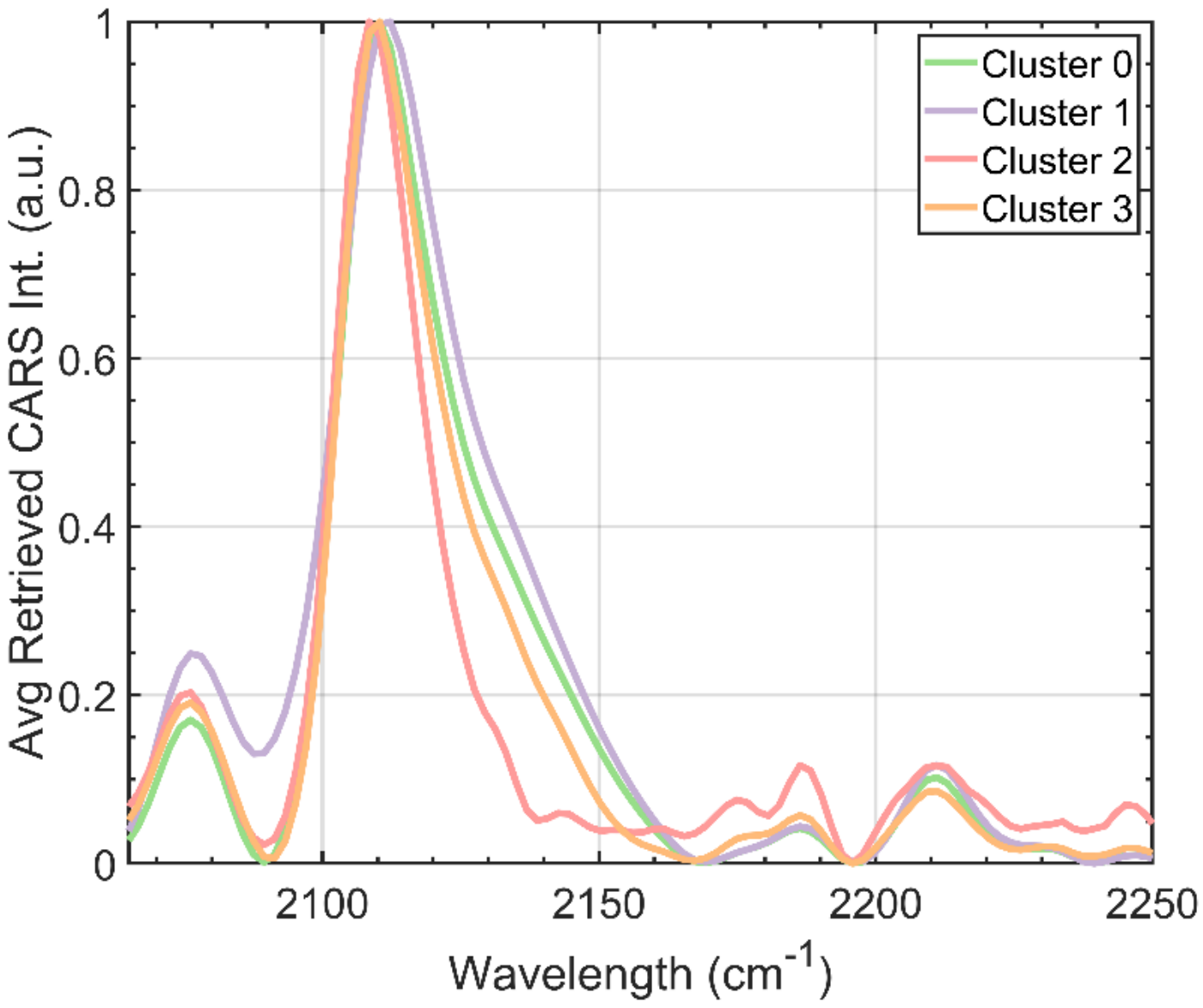

**Figure S4. Averaged spectrum of each cluster after UMAP.** The spectrum is normalized by the intensity of 2108 $cm^{-1}$.

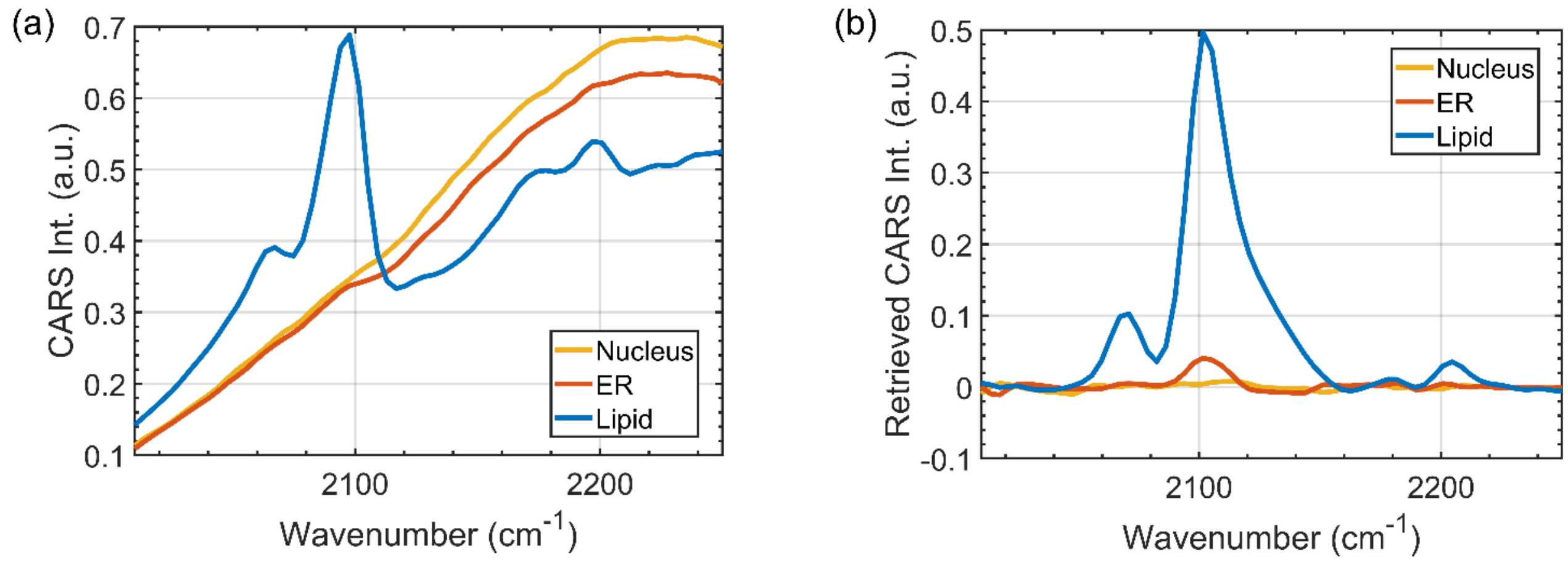

**Figure S5. Raw and retrieved spectrum of Nucleus, ER membrane and lipid by HELP-CARS imaging in the silent region.** (a) Raw spectrum. (b) Retrieved spectrum.

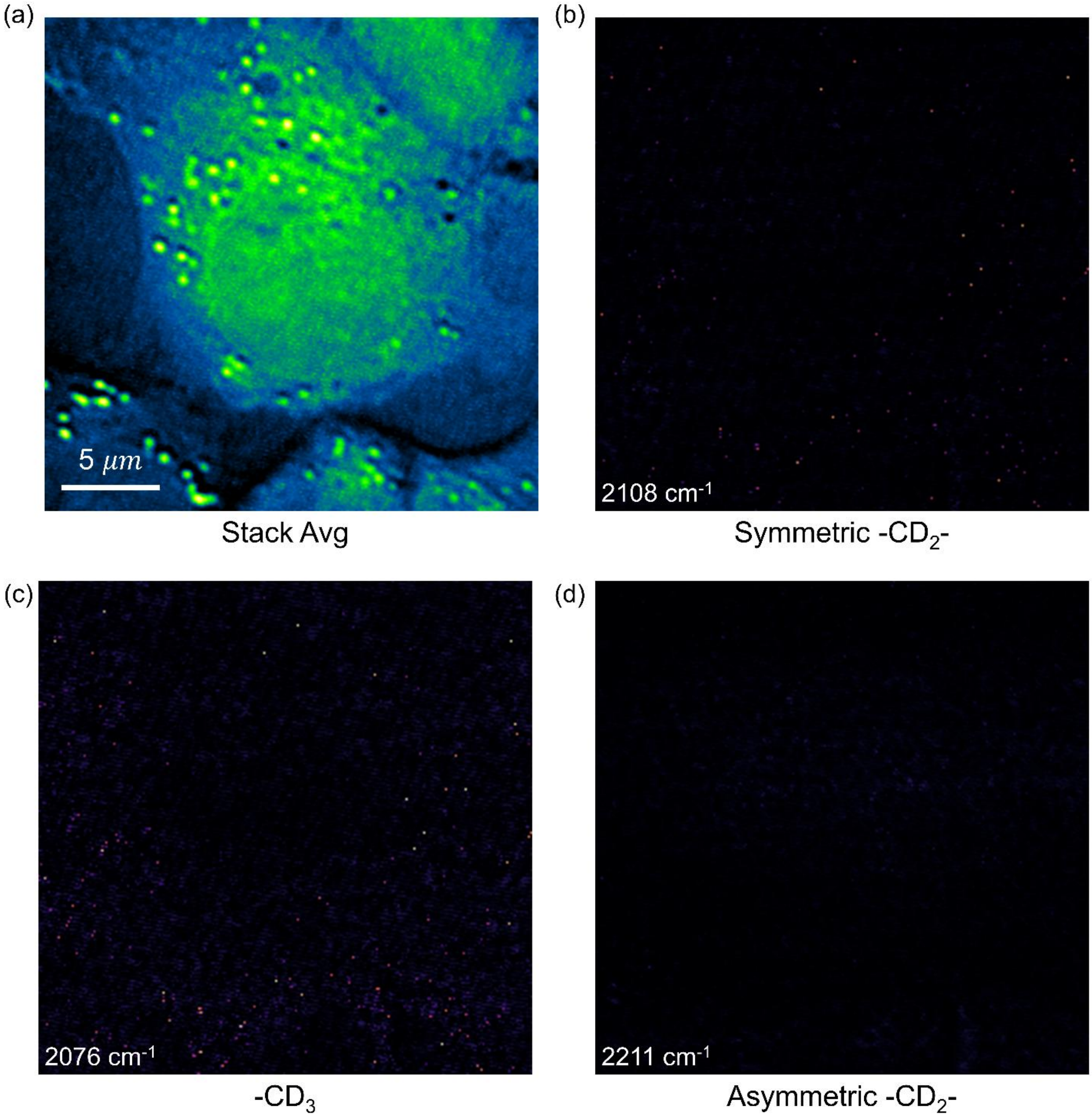

**Figure S6. Control Group of PA-d31 treated T24 cells by HELP-CARS in the silent region.** (a) Raw stack-averaged HELP-CARS image. (b) Raw symmetric $CD_2$ (2108 $cm^{-1}$) channel. (c) Raw $CD_3$ (2076 $cm^{-1}$) channel. (d) Raw asymmetric $CD_2$ (2211 $cm^{-1}$) channel.

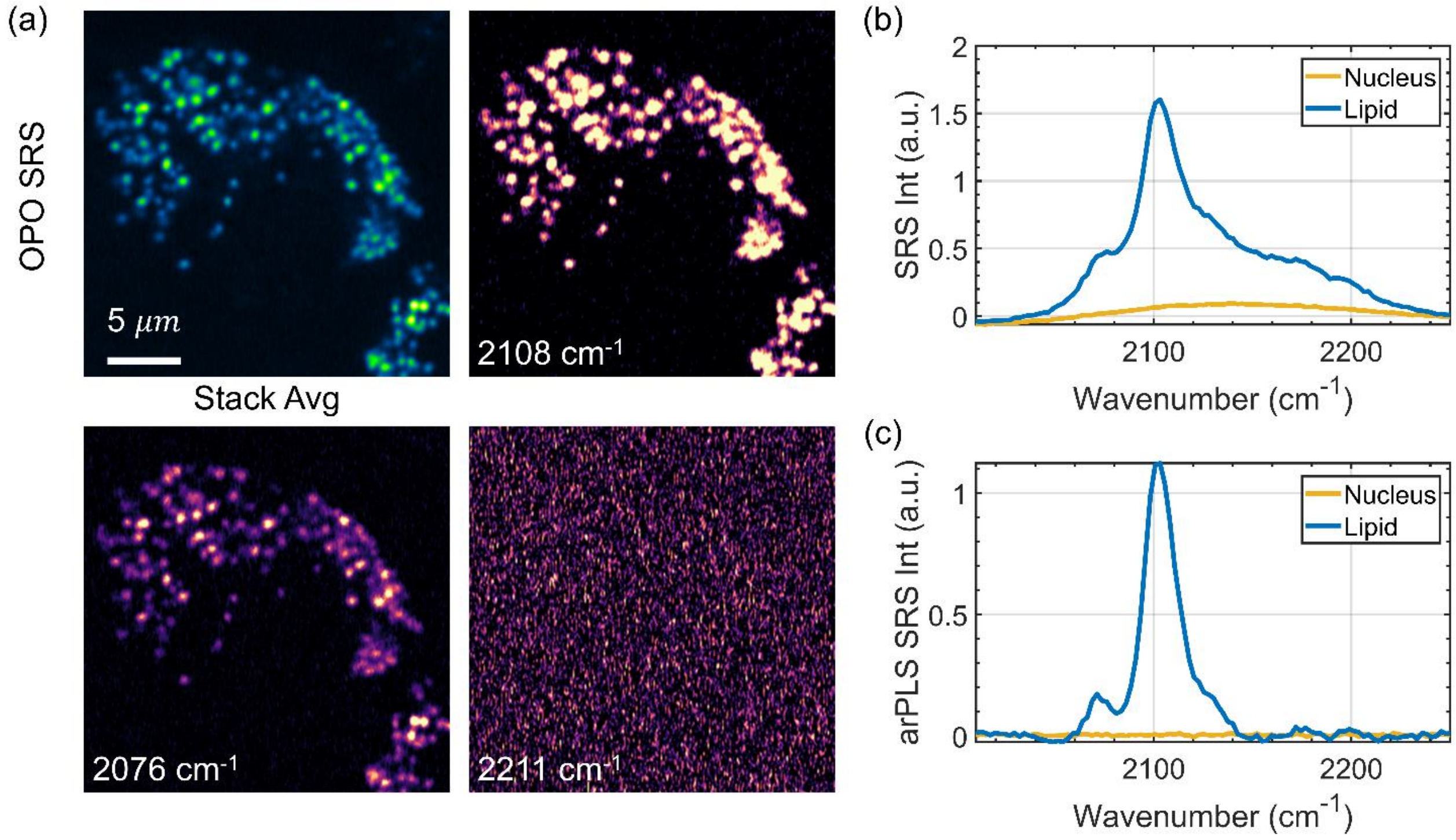

**Figure S7. SRS imaging of PA-d31 treated T24 cells in the silent region.** (a) hyperspectral OPO-SRS images of PA-d31 treated T24 cell. Top: stack-averaged image and symmetric $CD_2$ (2108 $cm^{-1}$) map. Bottom: $CD_3$ (2076 $cm^{-1}$) and asymmetric $CD_2$ (2211 $cm^{-1}$) maps. (b,c) Raw and arPLS-baseline-corrected OPO-SRS spectra from lipid droplets and nucleus.

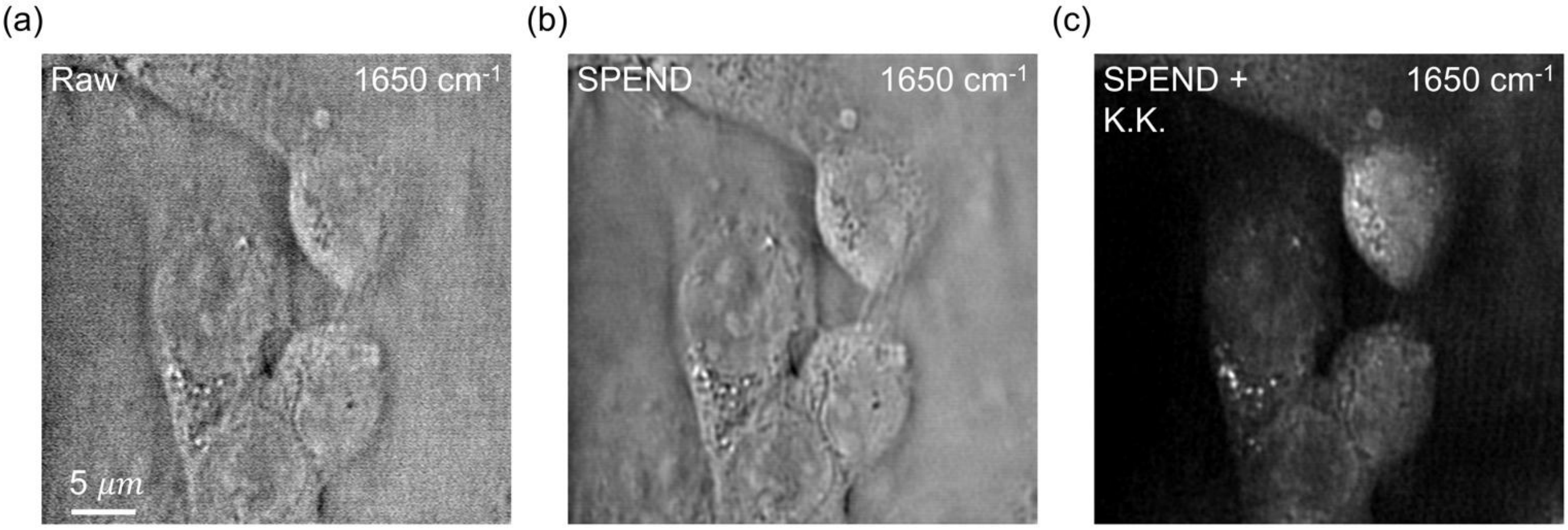

**Figure S8. M$\beta$CD treated T24 cells by HELP-CARS in the fingerprint region.** (a) Raw CARS image at 1650 $cm^{-1}$. (b) SPEND-denoised image at the same wavenumber. (c) SPEND followed by K.K. retrieval at the same wavenumber.

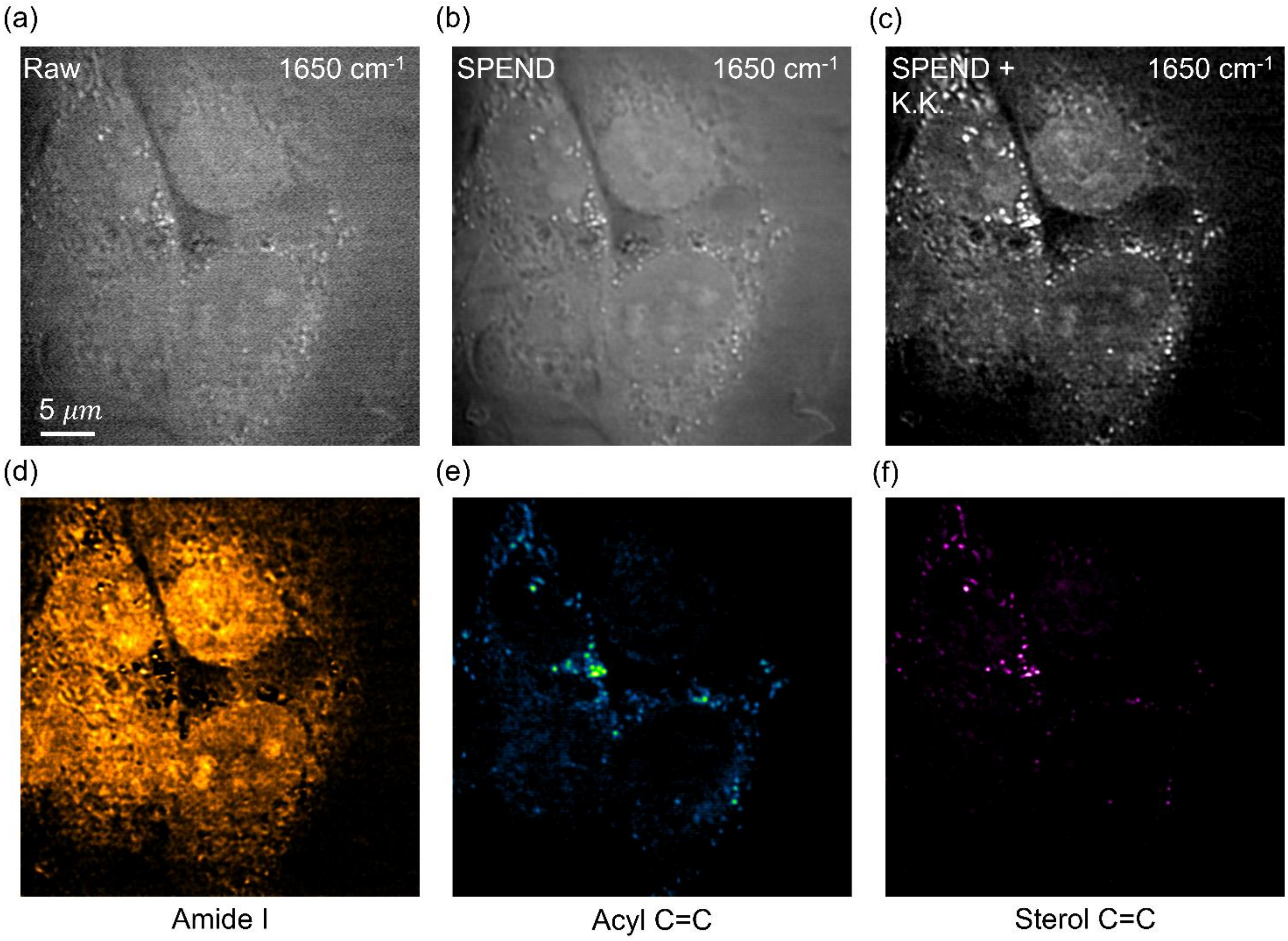

**Figure S9. Control group of T24 cells by HELP-CARS in the fingerprint region.** (a) Raw HELP-CARS image of control group at 1650 $cm^{-1}$. (b) SPEND-denoised image of control group at the same wavenumber. (c) SPEND followed by K.K. retrieval of control group at the same wavenumber. (d–f) Chemical maps derived from the KK-retrieved hyperspectral data: (d) Amide I band (protein-rich regions), (e) Acyl C=C band (lipid-rich regions), (f) Sterol C=C band (cholesterol-rich regions).